\documentclass[preprint,amsmath,amssymb]{revtex4}
\usepackage{amsmath}
\usepackage{graphicx}
\usepackage{amsfonts}
\usepackage{amssymb}
\usepackage{subfigure}
\usepackage{float}
\usepackage{color}

\begin{document}

\title{Modelling of strain fields in sheared colloidal glasses using Eshelby inclusions}
\author{Atharva Pandit$^{1}$, Antina Ghosh$^{1}$}
\author{Vijayakumar Chikkadi$^{1,2}$}
\thanks{past affiliation}
\affiliation{
$^1$ Physics Division, Indian Institute of Science Education and Research Pune, Pune 411008, India.\\
$^2$ Institute of Physics, University of Amsterdam, Science Park 904, 1098 XH Amsterdam, The Netherlands.\\}

\begin{abstract}
When amorphous solids are strained they display elastic deformation at small strain, however, beyond a critical strain they yield and begin to flow plastically. The origin of this plasticity lies in the irreversible rearrangement of particles. Such rearrangements have been shown to give rise to a long-ranged quadrupolar strain field, similar to Eshebly's spherical inclusions. However, their spatio-temporal organisation at finite temperatures and finite shear rate remains unclear. Here, we have investigated the strain field in sheared colloidal glasses. We show that the strain field in homogeneous flows can be modelled using a distribution of Eshelby inclusions. In particular, we show that the non-trivial decay of spatial strain correlations in sheared colloidal glasses is a result of elastic interactions between plastic rearrangements that form at spatially correlated locations.
\end{abstract}

\maketitle

\section{Introduction}
Amorphous solids are an important class of materials that appear in a variety of forms ranging from metallic glasses to polymeric glasses and soft materials made of emulsions, foams and granular matter \cite{barrat18,biroli11}. The mechanical properties of these solids differ significantly from each other, for example, the shear modulus of metallic glass is in the range of gigapascals and that of granular matter is a few Pascals. However, they display similar elastic and plastic properties. Their elasto-plastic deformation has attracted a lot of attention in the past decade and it has been a topic of intense research \cite{barrat18,barrat11}. Due to their disordered structure, the deformation of amorphous solids cannot be explained using the framework developed for crystal plasticity, which is based on dislocation motion \cite{falk98}. So, several questions relating to plasticity carriers in amorphous solids such as their characteristics, formation and their spatio-temporal organisation are central to our understanding of this topic.

In recent years, a new physics of deformation of amorphous solids is emerging due to insightful simulations and experiments. There is a developing consensus that plastic deformation in amorphous solids occur due to localized rearrangement of particles that give rise to long-range quadrupolar strain field. The strain field and the non-affine displacement around a plastic rearrangement resembles the Eshelby solution around a strained spherical inclusion in a homogeneous isotropic solid \cite{lemaitre,barrat06,schall07,chikkadi11}. The overall deformation occurs due to spatio-temporal interaction of such plastic rearrangements mediated by elasticity. This picture of deformation is true at low temperatures and low shear rates (close to quasi-static limit) \cite{finiteTGdot}. These features of amorphous plasticity have long been an integral part of many elasto-plastic models \cite{spaepen77,argon79,hebraud98,bocquet04}. Recent models have exploited these ideas to explain formation of shearbands and yielding in both two and three dimensions in quasi-static MD simulations of sheared amorphous solids \cite{ratul,zaccone17}. Further, mesoscopic simulations of amorphous plasticity based on lattice models have also incorporated the tensorial and long-range nature of elastic strain fields arising from plastic rearrangements \cite{zoe17}. Such studies have captured the complex patterning observed during plastic deformation under different loading conditions, and have also revealed a departure from mean-field depinning models.

On the other hand, the earliest attempt to directly observe localized plastic rearrangements in sheared amorphous systems in experiments was made using bidisperse bubble rafts \cite{argon82}. These experiments revealed the existence of two types of events - $1)$ diffuse plastic rearrangements involving $10-15$ bubbles undergoing complex internal rearrangement and $2)$ intense slip events of two adjacent nearly close packed bubble rows. However, the first evidence of plastic rearrangements interacting via long-range, quadrupolar strain field were obtained from shear experiments of colloidal glasses \cite{schall07,chikkadi11}. The robustness of these results were established by analysing spatial correlations of particle level strain and non-affine displacements. These experiments had also shed new light on shearbanding of colloidal glasses. When sheared at large strain rates, compared to inverse $\alpha-$relaxation time scale, colloidal glasses showed a transition from homogeneous flow to shearbanded flow \cite{chikkadi11}. The authors argued that at lower shear rates the plastic rearrangements occur everywhere in the system. However, at higher shear rates the plastic rearrangements align in the direction of shear leading to creation of weak zones and subsequently to shear banding. Recent investigations of quiescent glass forming liquids in three dimensions have also revealed the existence of long-range strain correlations that is characterized by $1/r^3$ decay and a quadrupolar structure \cite{varnik18}. The $1/r^3$ decay of strain correlations in $3D$ is consistent with the Eshelby solution for strain field around an inclusion \cite{eshelby}. However, in variance with these results, recent simulations of foam flows, close to the jamming transition, showed strain correlations decaying as $1/r^{\alpha}$ ($\alpha<2$) in two dimensions \cite{chikkadi15}. The slower decay of strain correlations was attributed to the superposition of strain fields of mutually interacting plastic rearrangements. This aspect of strain correlation has not been explored in experiments.

In this paper, we have investigated in detail the spatial correlations of local strain in sheared colloidal glasses in $3D$. We have shown that a simple model based on a distribution of Eshleby's spherical inclusions is sufficient to reproduce many features of strain field in experiments. In particular, we have shown that spatial strain correlations in our model capture the experimental observations when the spatial organisation of inclusions is strongly correlated. Our results highlight an intricate interplay of disorder and elasticity, and they offer new insights into flow of colloidal glasses and amorphous solids in general.

\section{Experimental methods}

\subsection{Sample preparation and shear measurement}
We prepared a colloidal glass by suspending sterically stabilized fluorescent polymethylmethacrylate particles in a density and refractive index matching mixture of cycloheptyl bromide and cis-decalin. The particles have a diameter of $\sigma = 1.3 \mu m$, and a polydisperity of $7\%$ to prevent crystallization. The suspension was centrifuged at an elevated temperature to obtain a dense sediment, which was subsequently diluted to get a suspension of desired volume fraction $\phi\sim0.60$. The sample was sheared using a shear cell that had two parallel boundaries separated by a distance of $\sim 50\sigma$ along the $z-$direction. A piezoelectric device was used to move the top boundary in the $x-$direction to apply shear rates in the range $10^{-5}-10^{-4}s^{-1}$. To prevent boundary-induced crystallization in our samples, the boundaries were coated with a layer of polydisperse particles. Confocal microscopy was used to image the individual particles and to determine their positions in three dimensions with an accuracy of $0.03 \mu m$ in the horizontal and $0.05 \mu m$ in the vertical direction. We tracked the motion of $\sim 2 \times 10^{5}$ particles during a $25$-min time interval by acquiring image stacks every $60~s$. All the measurements presented here were made in the steady state, after the sample had been strained to $100\%$ at shear rate of $1.5\times 10^{-5}s^{-1}$, as confirmed by other independent rheological measurements.

\subsection{Spatial strain correlations}
We study correlations in the deformation of the glass by decomposing the particle motion into affine and non-affine components. To compute these quantities we follow particle trajectories and identify the nearest neighbors of each particle as those separated by less than $r_{0}$, the first minimum of the pair correlation function. We subsequently determine the best affine deformation tensor ${\bf\Gamma}$ describing the transformation of the nearest neighbor vectors, ${\bf d_i}$, over the strain interval $\delta \gamma$ \cite{falk98}, by minimizing $D^2_{min} = (1/n) {\sum_{i=1}^{n}}({\bf d_i}(\gamma + \delta \gamma) - {\bf \Gamma}{\bf d_i}(\gamma))^2$. The symmetric part of ${\bf \Gamma}$ is the local strain tensor. The remaining non-affine component $D^2_{min}$ has been used as a measure of plastic deformation \cite{falk98}. We will investigate the correlations in the fluctuations of the principal shear strain component $\epsilon_{xz}$ using the following expression \cite{chikkadi11}:
\begin{equation}
C_{\epsilon}({\bf \delta r}) = \frac{ \left< \epsilon_{xz}({\bf r + \delta r}) A({\bf r})
\right> - \left< \epsilon_{xz}({\bf r}) \right> ^{2} } { \left< \epsilon_{xz}({\bf r})^{2}
\right> - \left< \epsilon_{xz}({\bf r}) \right> ^{2} }  ,
\label{c_r}
\end{equation}
where angular brackets denote average over all the particles in the systems and several strain steps. $C_{\epsilon}$ correlates values of $\epsilon_{xz}$ at locations separated by ${\bf \delta r}$, this way we capture the elastic response of the system in the steady state.

\section{Results}
\subsection{Spatial correlations of strain in sheared colloidal glasses}

\begin{figure}[htp]
\centering
\begin{tabular}{ccc}
\includegraphics[width=.4\textwidth]{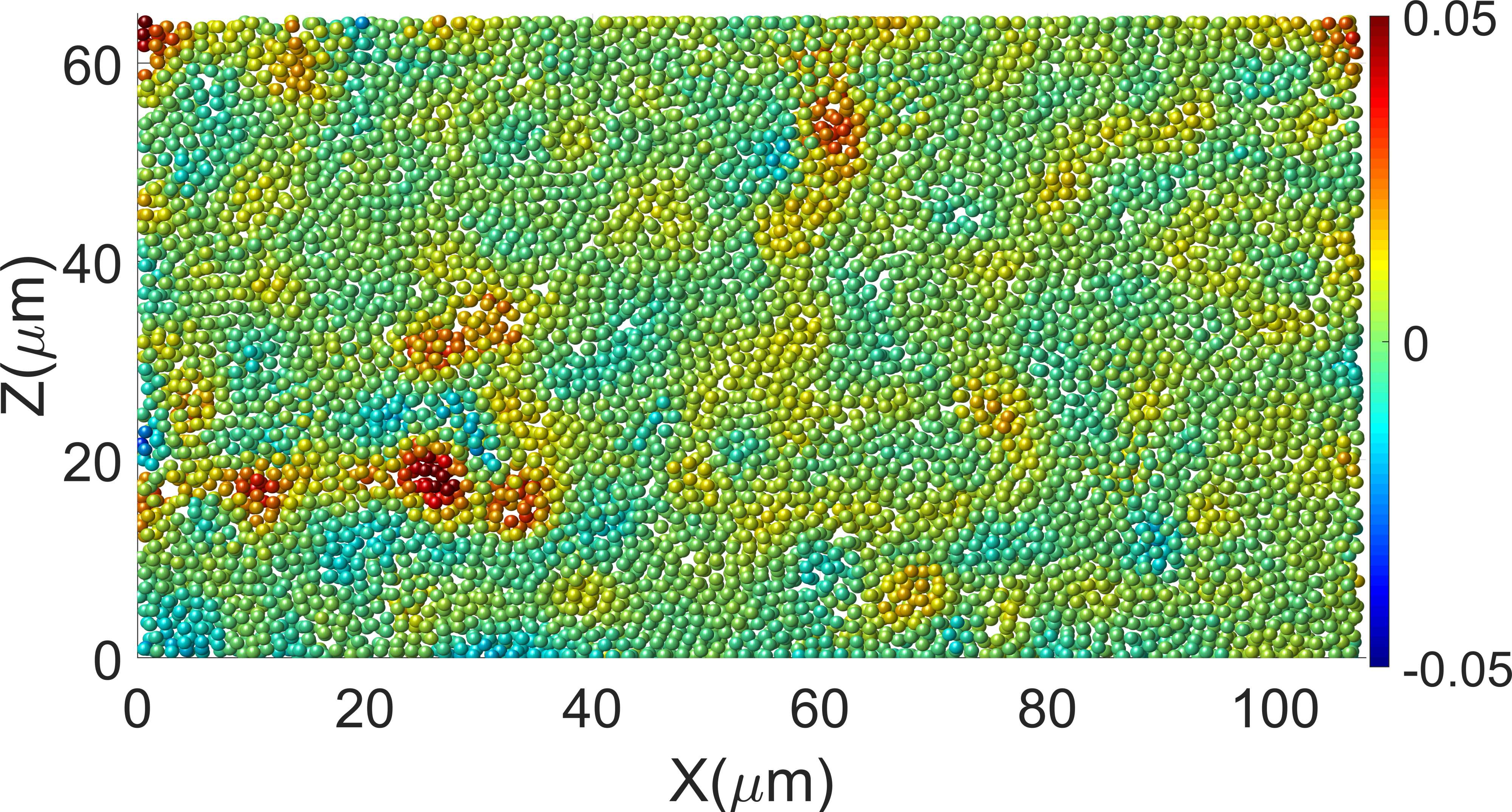} &
\includegraphics[width=.25\textwidth]{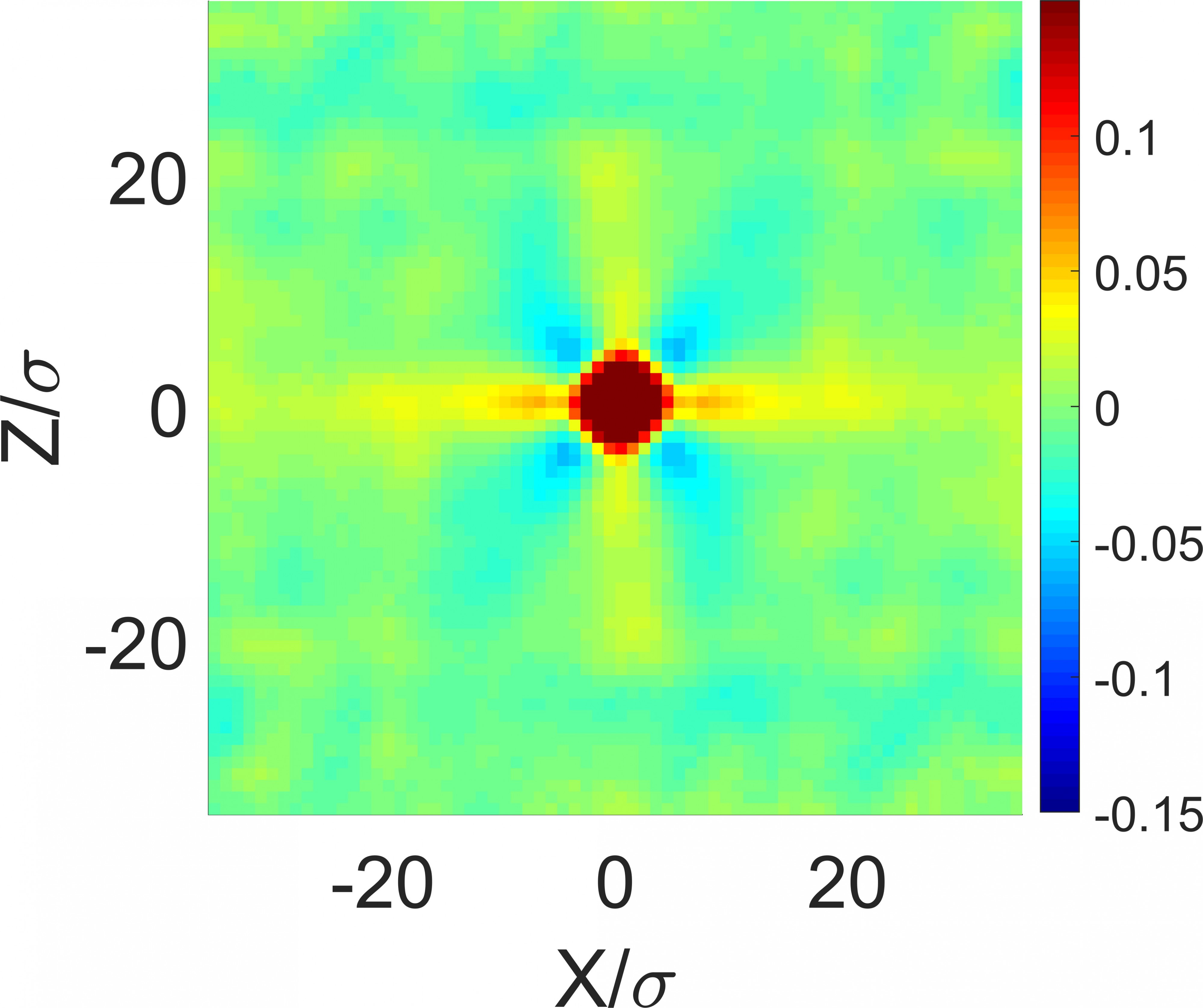} &
\includegraphics[width=.25\textwidth]{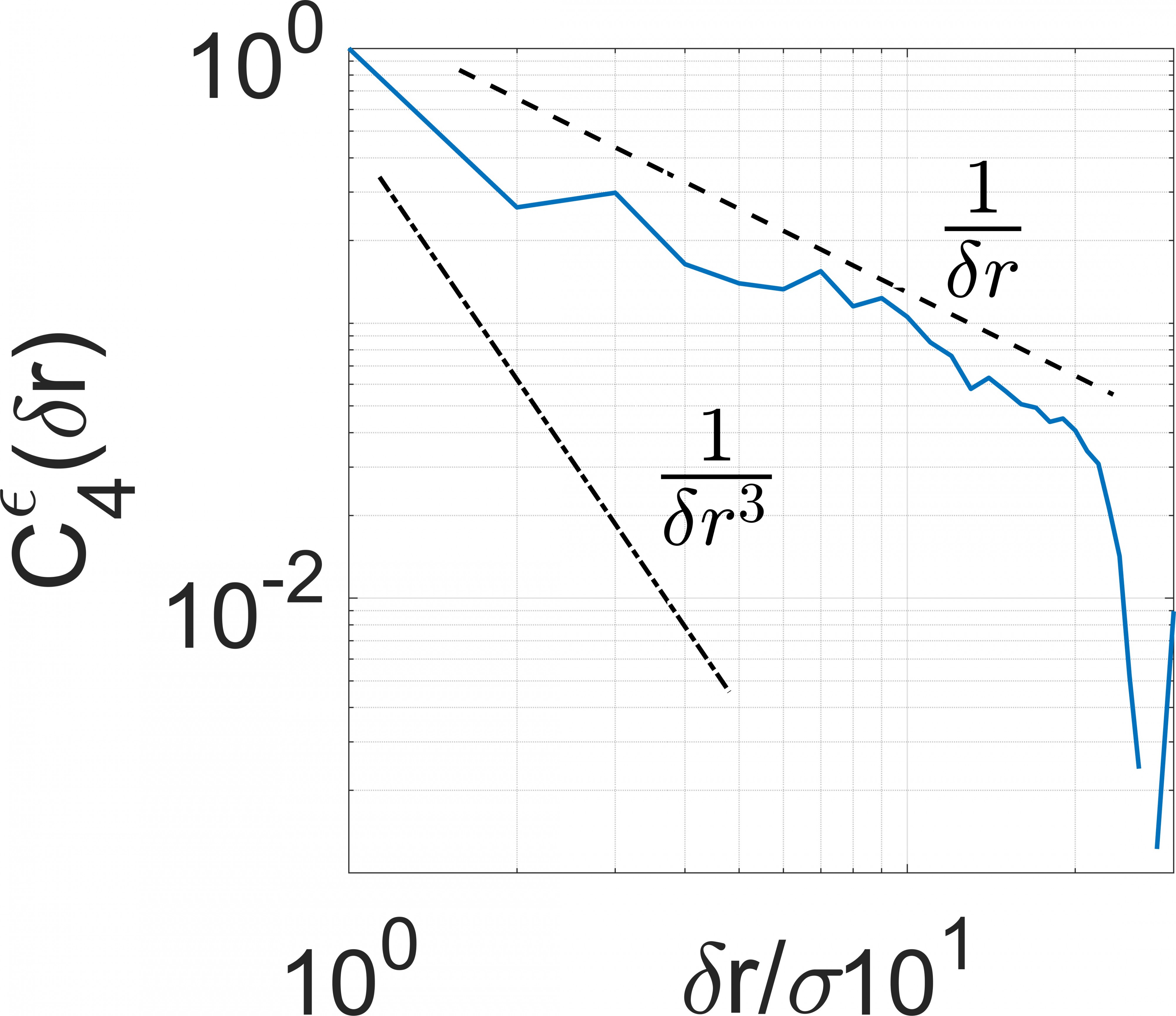} \\
\textbf{(a)}  & \textbf{(b)} & \textbf{(c)}  \\[6pt]
\end{tabular}
\caption{(a) Reconstruction of local shear strain ($\epsilon_{xz}$) in a colloidal glass sheared at $\dot{\gamma} = 1.5 \times 10^{-5} s^{-1}$. For ease of visualisation particles in a $5\mu m$ thick region along $y$ direction are shown. Particles are color coded to indicate their value of $\epsilon_{xz}$. (b) Spatial correlations of $\epsilon_{xz}$ in the shear plane $xz$ for $\delta y=0$ is shown. The quadrupolar symmetry of strain correlations is very clear from this figure. (c) The projection of strain correlations $C_{\epsilon}^{4}(\delta r)=\int_{0}^{2\pi} C_{\epsilon}(\delta x,0,\delta z)~cos(4\theta)~d\theta$ is shown in the figure, along with curves that decay as $1/\delta r$ (dash line) and $1/\delta r^3$ (dash-dot line).}
\label{fig_1}
\end{figure}

We have investigated strain correlations of colloidal glasses in a steady state flow. As reported earlier, our experiments reveal a transition from homogeneous flow to heterogeneous flow (shearbanding) beyond a critical strain rate of $1\times10^{-4}s^{-1}$ \cite{chikkadi11}. The current analysis is performed at a shear rate of $1.5\times10^{-5}s^{-1}$ where the systems flows homogeneously in the steady state. The relaxation time of the system was determined by measuring the mean square displacement of particles in quiescent conditions, and it was found to be $1\times10^{4}s$. We determine local shear strain of particles from the relative motion with respect to its nearest neighbors over a strain interval $\delta\gamma\sim0.036$. A spatial reconstruction of microscopic shear strain $\epsilon_{xz}$ is shown in Fig.~1a. Red and blue regions indicate zones where large shear strain in the system are localised. We determine the elastic response of the system by computing spatial correlations of shear strain, $C_{\epsilon}(\bf \delta r)$. At the short strain intervals considered here, the correlation functions do not depend significantly on the specific strain interval chosen and appear robust. Correlations in the $x$-$z$ plane are obtained by taking ${\bf \delta r} = (\delta x,0,\delta z)$; a corresponding color coded representation of the correlation function $C_{\epsilon}(\bf \delta r)$ is shown in Fig.~1b. The correlation function displays a four-fold symmetry that is characteristic of a quadrupolar strain field around an Eshelby inclusion \cite{chikkadi11}. This suggests that high strain regions (blue and red zones in Fig.1a) act as strained inclusions in an elastic matrix. Further, we project the correlation function on to its corresponding circular harmonic, using $C_{\epsilon}^{4}(\delta r)=\int_{0}^{2\pi} C_{\epsilon}(\delta x,0,\delta z)~cos(4\theta)~d\theta$. In Fig.1c we show $C_{\epsilon}^{4}(\delta r)$ of the strain correlation shown in Fig.1b. Similar to non-affine displacements \cite{chikkadi11}, the spatial strain correlations display an algebraic decay $1/r^{\alpha}$, where $\alpha\sim1$. However, this decay is slower than the $1/r^{3}$ behavior predicted by Eshelby for a strained spherical inclusion in homogeneous isotropic elastic solid \cite{eshelby}. Apparently, the deviation seems to stem from the fact that strain correlations are capturing the elastic response of not a single inclusion but several interacting inclusions. To verify this idea we will follow the method outlined in \cite{chikkadi15}, which is described in the following section.

\subsection{Spatial organisation of Eshelby inclusions}

Recent studies of amorphous plasticity have used a minimal model based on energetics of interacting Eshelby inclusions to explain yielding and shearbanding in quasi-static simulations of sheared amorphous solids \cite{ratul,zaccone17}. In addition, the slow decay of strain correlation in simulations of foam flows was also modeled using a distribution of Eshelby inclusions \cite{chikkadi15}. Motivated by these studies we model the strain fields in our experiments using Eshelby inclusions and study their correlations to elucidate the slow decay of strain correlations that is seen in Fig.1c. The underlying idea is to generate synthetic strain fields using Eshelby inclusions, and to compute their strain correlations. It is evident from Fig.1a that such heterogeneous strain fields should result from multiple inclusions, and not a single inclusion.


\begin{figure}[htp]
\centering
\begin{tabular}{ccc}
\includegraphics[width=.37\textwidth]{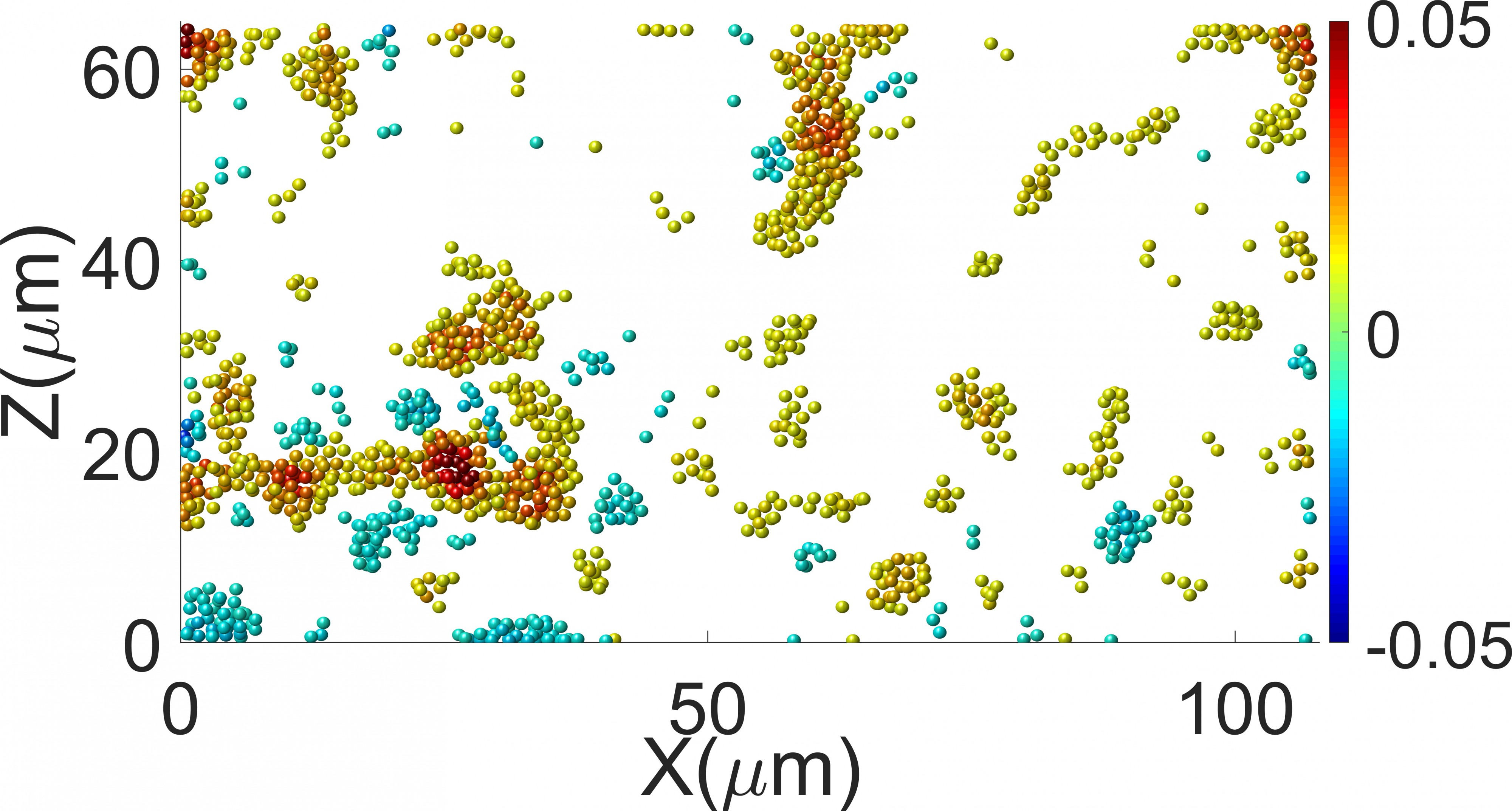} &
\includegraphics[width=.37\textwidth]{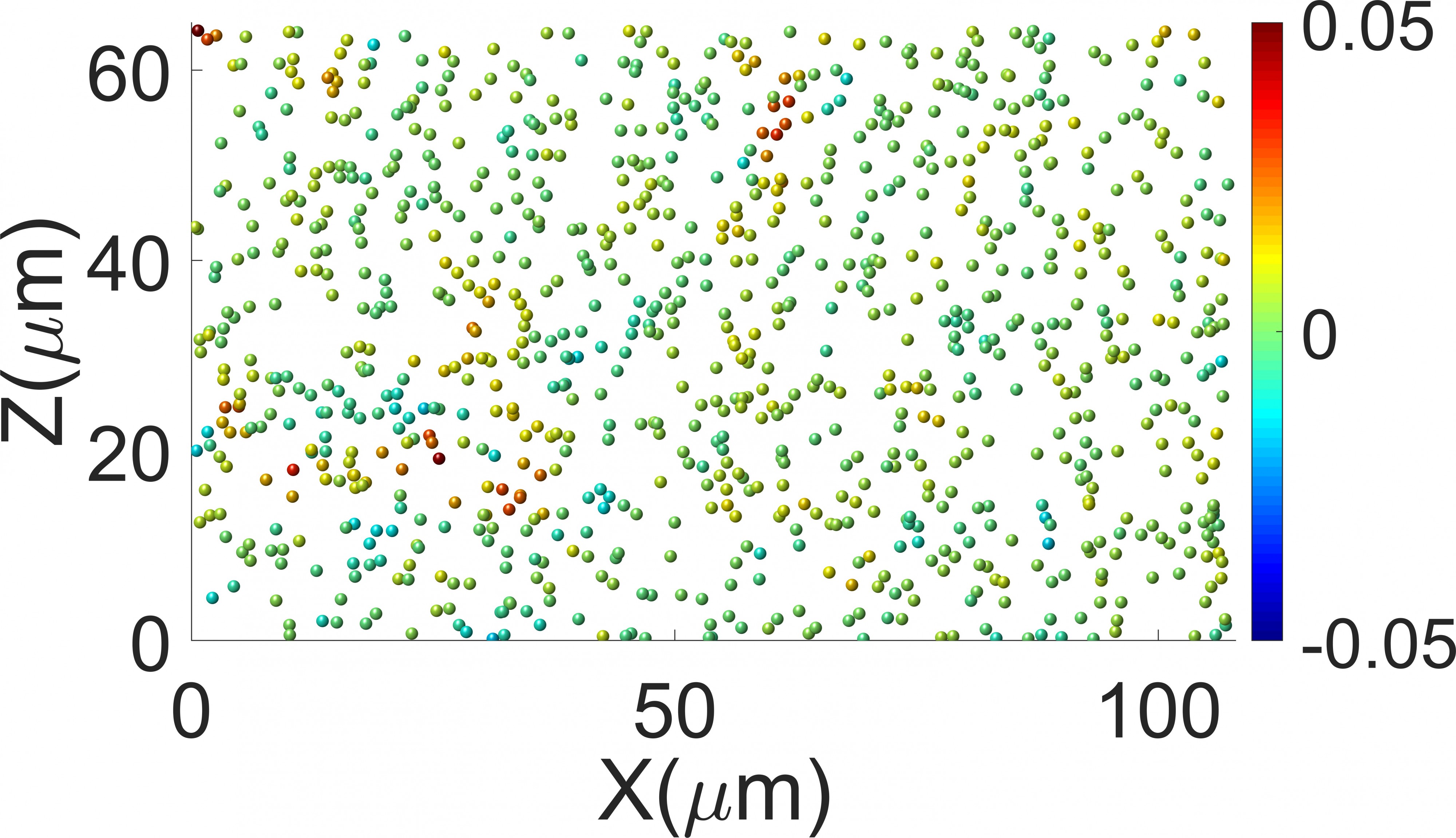} &
\includegraphics[width=.23\textwidth]{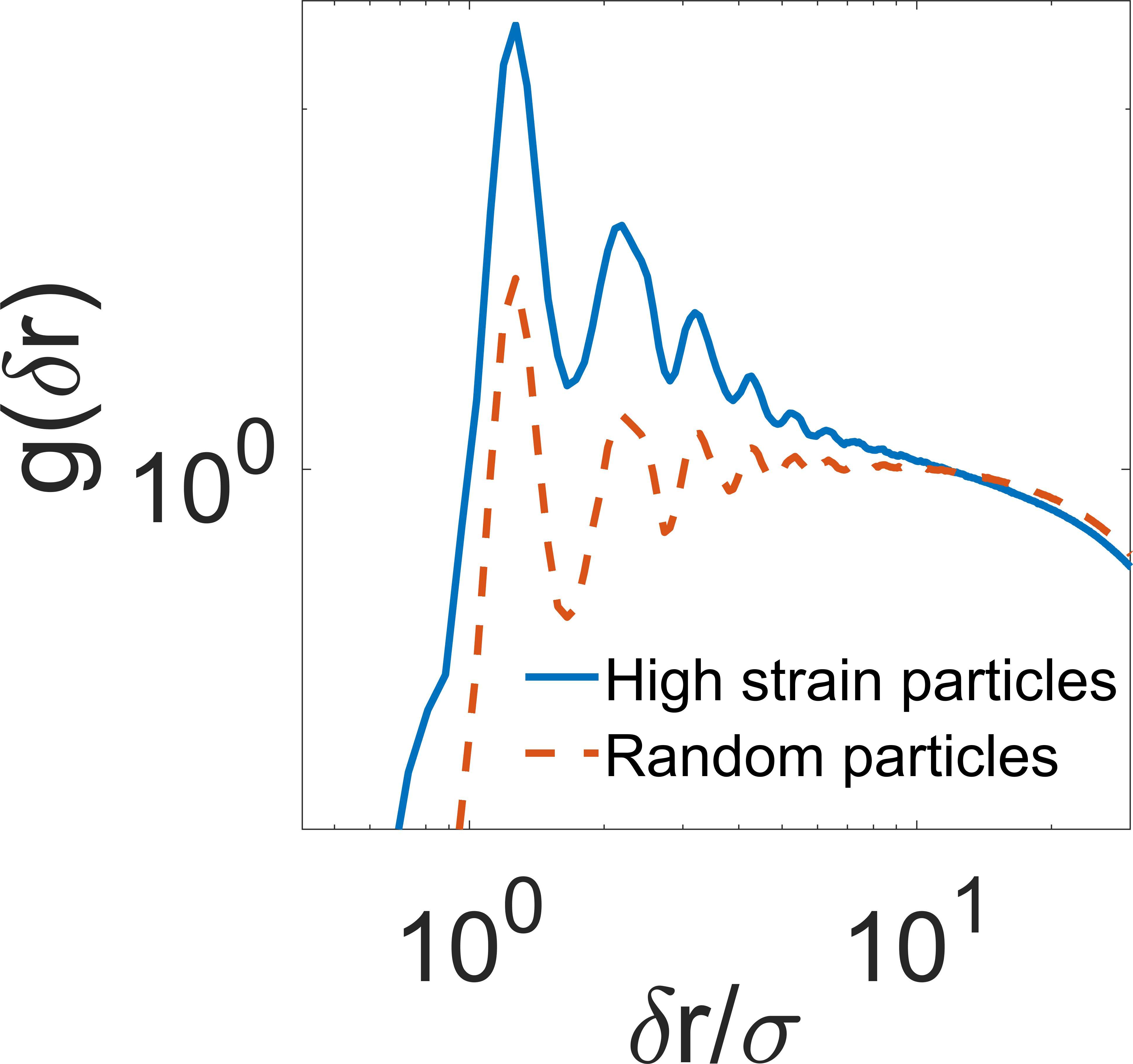} \\
\textbf{(a)}  & \textbf{(b)} & \textbf{(c)}  \\[6pt]
\end{tabular}
\caption{(a) Spatial distribution of particles with shear strain $\epsilon_{xz}>0.95\epsilon_{xz}^{max}$, where $\epsilon_{xz}^{max}$ is the maximum strain of particles. These are considered as inclusion centers in the first case. (b) Inclusion centers are obtained from random selection of particles. Note that the particles in (a) \& (b) are once again shown in a $5\mu m$ thick region in $y-$direction. (c) Normalized pair correlation function of inclusion centers. The continuous line corresponds to inclusion centers obtained from particles with top $5\%$ strain and dashed line is obtained for random selection of inclusion centers.}
\label{fig_2}
\end{figure}

The natural candidates as inclusion cores are the high strain particles that appear as red and blue particles in Fig.~1a. However, to investigate the effect of spatial organisation of inclusions, we will follow two different ways of selecting inclusions : a) the top $5\%$ strained particles in our systems are taken as inclusion cores, and b) particles are chosen at random with their strain values as inclusion centers. It is assumed that the size of inclusion core is about a particle diameter \cite{jensen14}. A representative spatial distribution of particles that were selected based on the above two criteria is shown in Figs.~2(a) \& 2(b). It appears that the inclusion centers are clustered and are more correlated in Fig.~2(a) in comparison to inclusions in Fig.~2(b). To highlight these differences we plot the pair correlation function of inclusion centers using the following expression
\begin{equation}
g(r)=\frac{V}{4 \pi r^{2} N^{2}}\left\langle\sum_{i} \sum_{j \neq i} \delta\left(r-r_{i j}\right)\right\rangle.
\end{equation}
The angular brackets in the above equation indicates that the pair correlation function is averaged over all inclusions in the system and several strain steps in the steady state is shown in Fig.~2(c). Clearly, the short-range correlations are stronger when particles with top $5\%$ strain are considered as inclusion centers, whereas the inclusion centers corresponding to random selection appear to be weakly correlated.

\subsection{Synthetic strain field using Eshelby inclusions}


\begin{figure}[htp]
\centering
\begin{tabular}{cc}
\includegraphics[width=.43\textwidth]{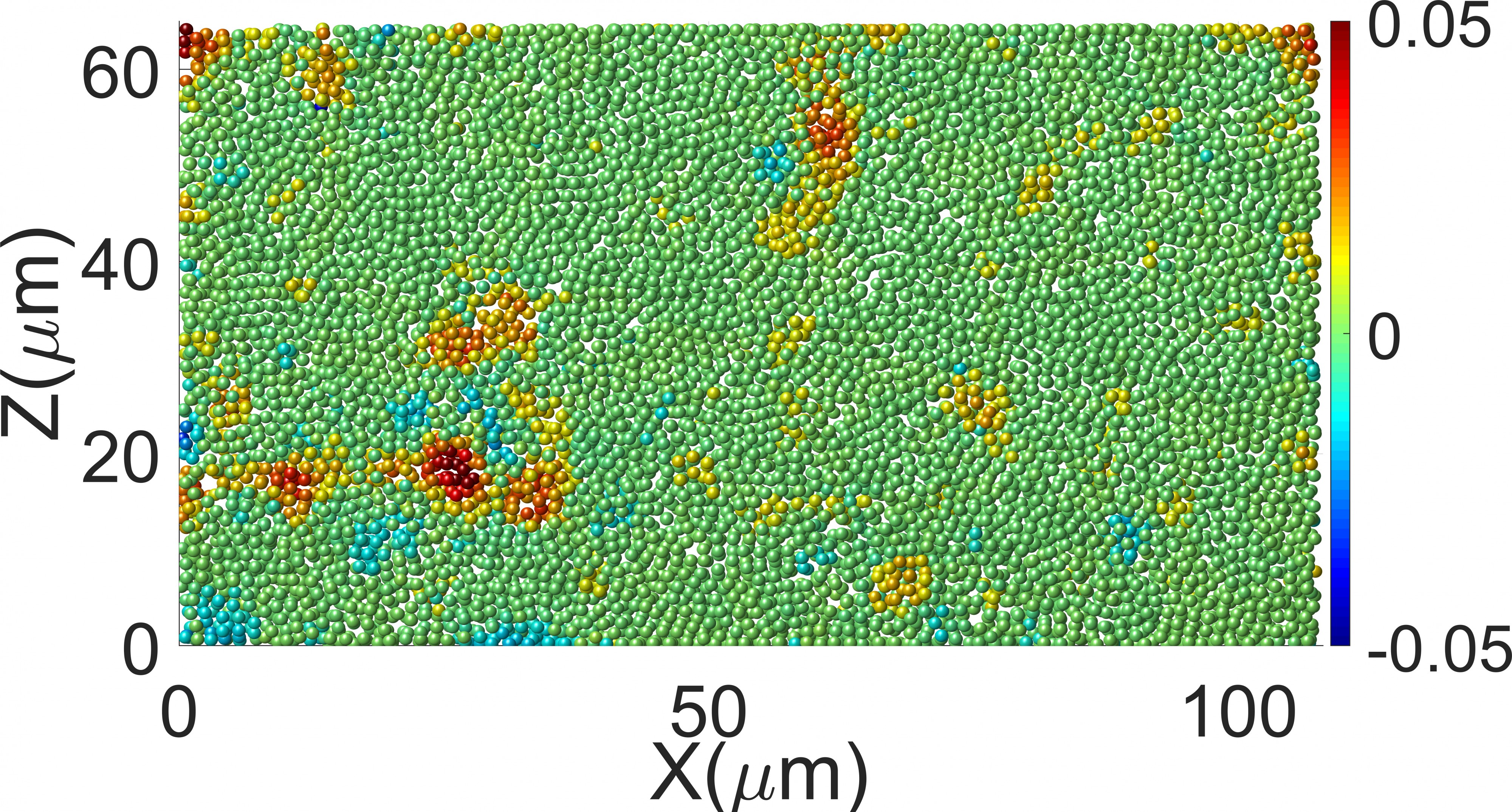} &
\includegraphics[width=.44\textwidth]{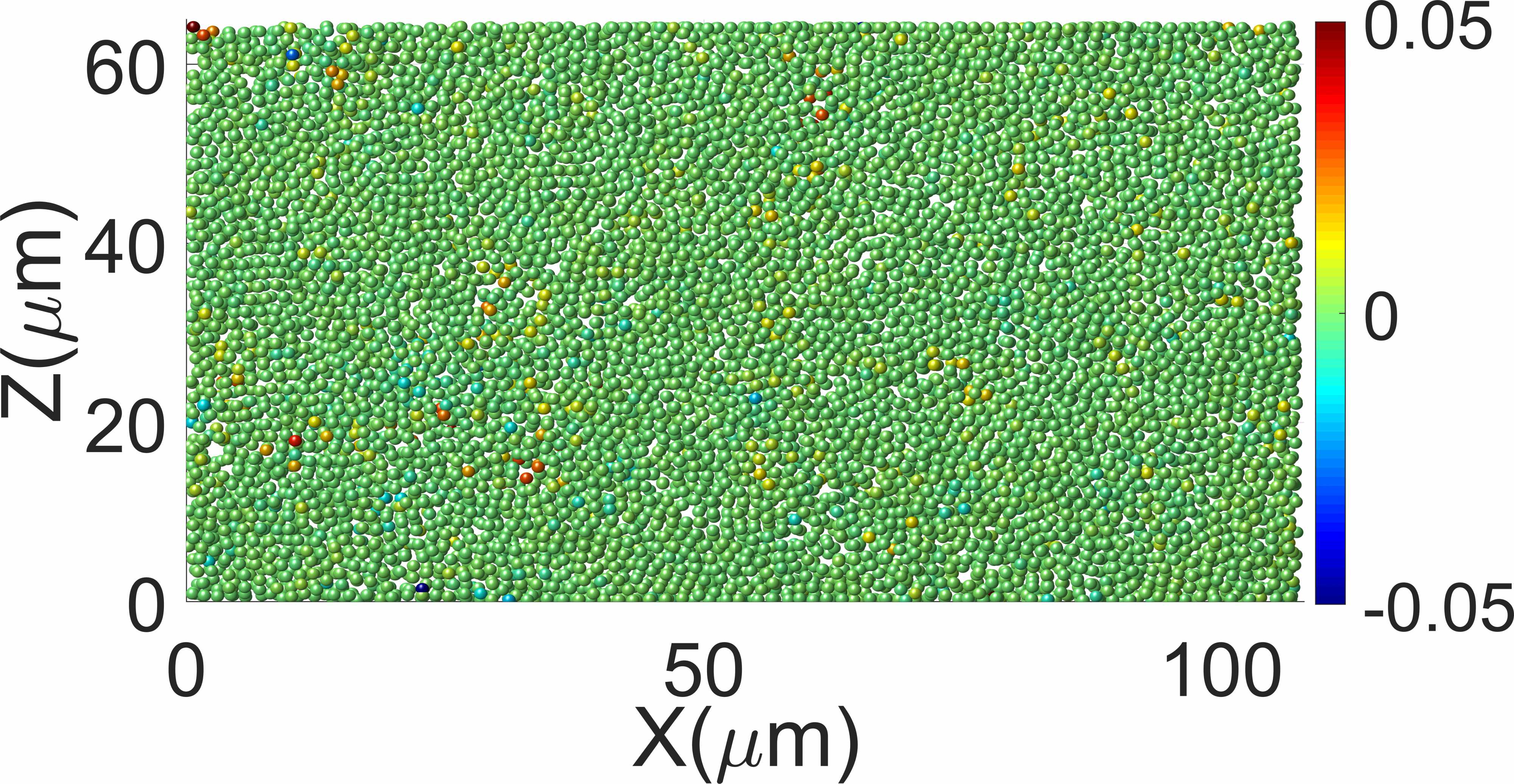} \\
\textbf{(a)}  & \textbf{(b)} \\[6pt]
\end{tabular}
\medskip
\begin{tabular}{ccc}
\includegraphics[width=.31\textwidth]{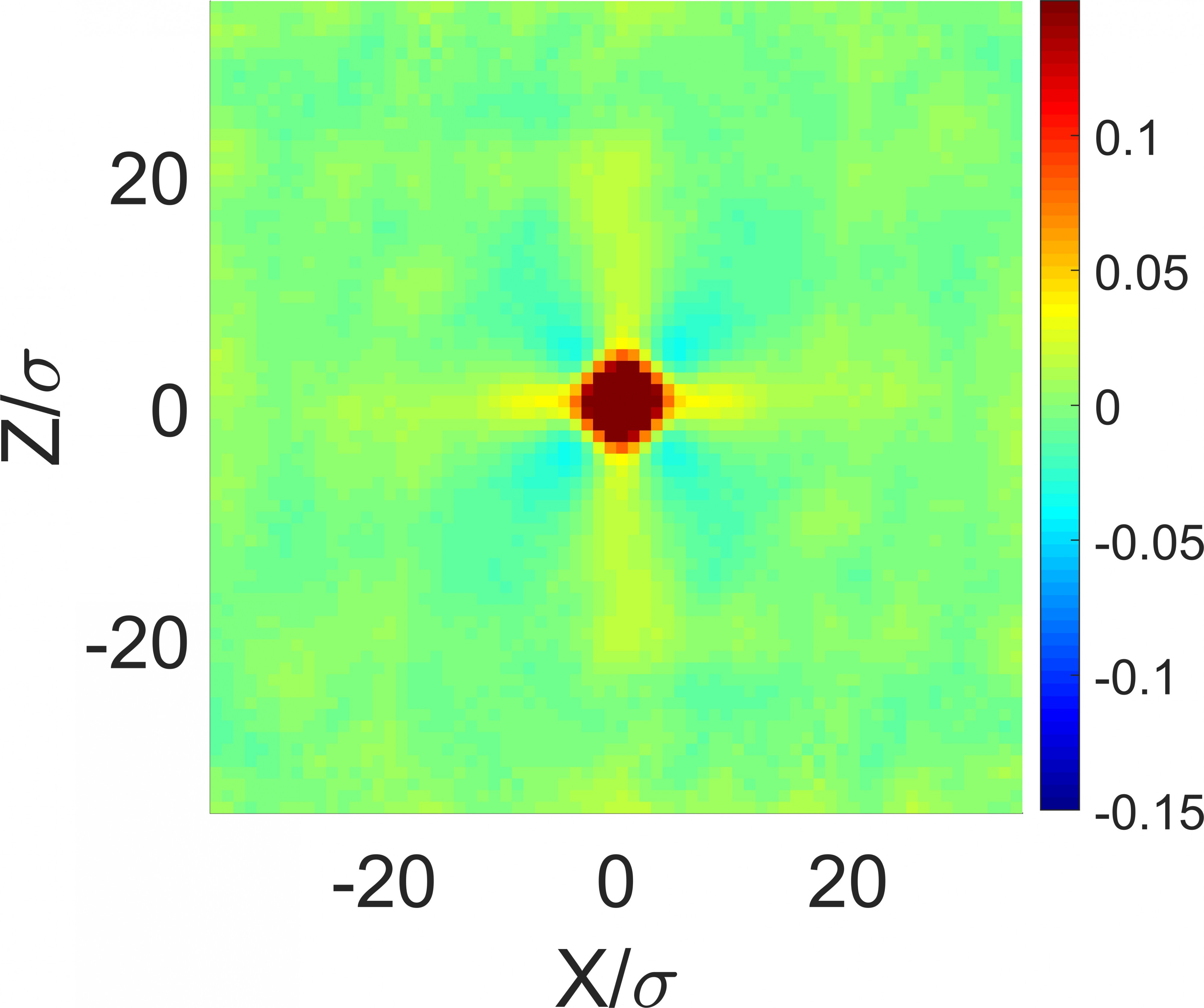} &
\includegraphics[width=.31\textwidth]{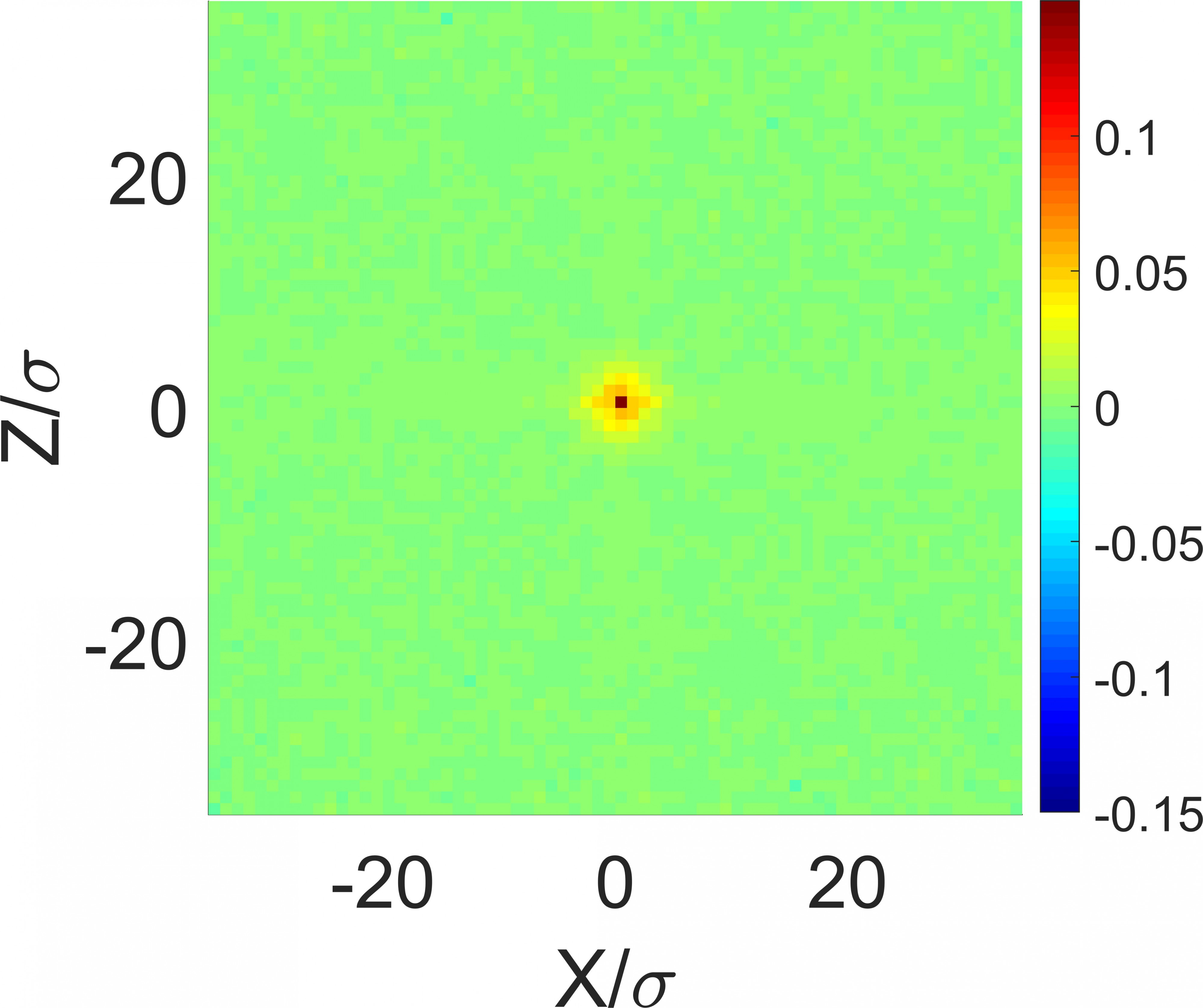} &
\includegraphics[width=.28\textwidth]{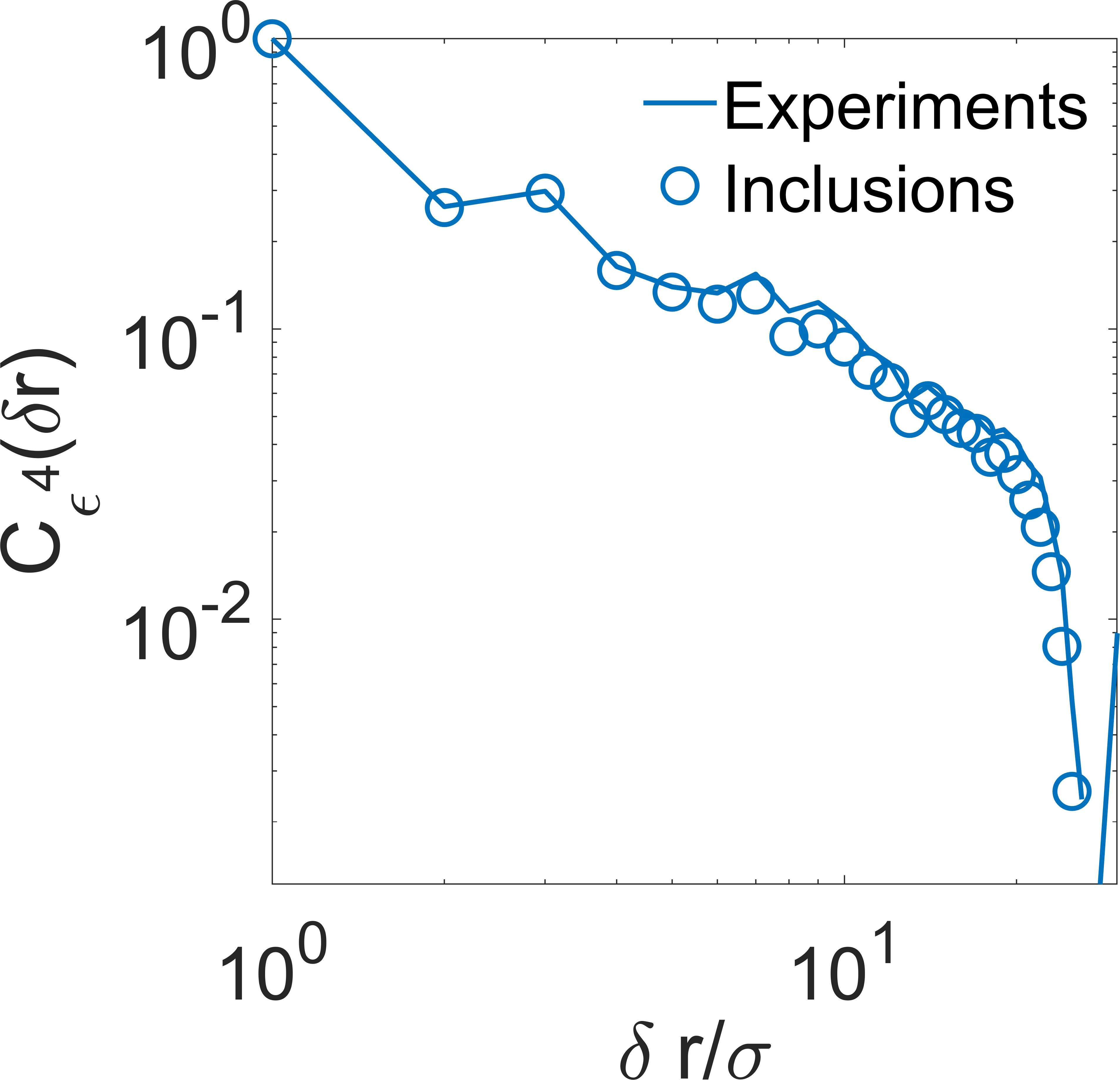} \\
\textbf{(c)}  & \textbf{(d)}  & \textbf{(e)} \\[6pt]
\end{tabular}
\caption{The figures in the top panel refer to shear strain $\epsilon_{xz}$ of particles that is obtained from the superposition of strain due to all inclusion - $\epsilon_{xy}^{i}=\sum_{j=1,j\neq i}^{N_{inc}}\varepsilon_{xz}^{I}(r_ij,\theta_{ij})$, where $r_{ij}$ is the distance of particle $i$ from an inclusion $j$, and $\theta_{ij}= cos^{-1}(z_{ij},r_{ij})$. (a) Particles with shear strain $\epsilon_{xz}>0.95*\epsilon_{xz}^{max}$, were considered as inclusion centers. (b) Particles were randomly selected as inclusion centers. Spatial correlations of synthetic strain fields when top $5\%$ strained particles are considered as inclusion centers (c), and when inclusion centers are selected randomly (d).(e) Projections of strain correlations $C_{4}^{\epsilon}(\delta r)$ obtained from superposition of inclusions at high strain particles (circles) is compared with experimental results (thick curve).}
\label{fig_3}
\end{figure}

We now turn to synthetic strain fields where the strain on a particle is the superposition of strain fields due to multiple Eshelby inclusions. The elastic shear strain field around a single inclusion $\varepsilon_{xz}^{I}$ in an isotropic homogeneous elastic solid is given by the following expression \cite{jensen14}

\begin{multline}
\varepsilon_{xz}^{I}(r,\theta,y)= \frac{a^{3}\epsilon_{0}} {4\left(r^{2}+y^{2}\right)^{9/2}}
\left\{ 9 a^{2} c r^{4}-(2+3 c) r^{6} + 9 c r^{2} \left( -8 a^{2}+5 r^{2} \right) y^{2} + 6 \left[ r^{2}+4 c\left(a^{2}+r^{2}\right)\right] y^{4} \right. \\
\left. +4(1-6 c) y^{6} + 15 c r^{4} \left[ -7 a^{2} + 5 \left( r^{2}+y^{2} \right) \right] \cos(4\theta) \right\}.
\end{multline}
When $y = 0$, the expression reduces to

\begin{equation}
\varepsilon_{xz}^{I}(r,\theta)=
\begin{cases}
\epsilon^{0},r<a\\
\frac{\epsilon_{0}a^3}{4r^5}\left[9a^2c-(2+3c)r^{2}-15c(7a^2-5r^2)cos(4\theta)\right],r\geq~a,\\
\end{cases}
\end{equation}

where $a$ is the core size of inclusion, $\epsilon_0$ is the core strain, $c=1/4(4-5\nu)$ is a dimensionless elastic constant, with $\nu$ Poisson's ratio, $\bf{r}(x,z)$ the position vector with origin at the center of the inclusion, and $\theta=cos^{-1}(z,r)$. Details of the derivation are provided in \cite{jensen14}. The core strain of an inclusion $\epsilon_0$ is the shear strain of the particle obtained from experiments. The shear strain on any particle in the system is a superposition of shear strains due to all inclusions $\epsilon_{xy}^{i}=\sum_{j=1,j\neq i}^{N_{inc}}\varepsilon_{xz}^{I}(r_{ij},y_{ij},\theta_{ij})$,
where $r_{ij}$ is the distance of particle $i$ from an inclusion $j$, and $\theta_{ij}= cos^{-1}(z_{ij},r_{ij})$. The synthetic strain fields obtained this way are shown in Figs.~3(a) \& 3(b). The reconstructions of shear strains shown in Fig.~3(a) is obtained from superposition of strain due to inclusions at top strained particles, whereas Fig.~3(b) is obtained when particles are selected randomly as inclusion centers. For ease of comparison, the strain map in Figs.~3(a) \& 3(b) display the same set of particles that are shown in Fig.~1(a). It is apparent from these figures that the strain map in Figs.~3(a) compares well with experiments in Fig.~1(a). To establish this further, we have computed spatial correlations of shear strain and their projections $C_{\epsilon}^{4}(\delta r)$ in Figs. 3(c)-3(e). From Figs.~3(c) \& 3(d) it is clear that the quadrupolar symmetry persists only when the inclusion centers are clustered and spatially correlated, however, it disappears when the inclusion centers are randomly located. Further, when we plot the projections of strain correlation function $C_{\epsilon}^{4}(\delta r)$  for inclusions at high strain particles, along with the experimental results, we see an excellent agreement. These results establish that strain correlation for sheared systems show deviation from the $1/r^3$ behavior predicted by Eshelby solution due to strongly correlated formation of inclusions.

\section{Conclusions}

We have investigated strain fields and strain correlations in a sheared colloidal glass. The quadrupolar projections of strain correlations reveal that they decay as $1/r^{\alpha}$ ($\alpha\sim1$) instead of $1/r^{3}$ as predicted by Eshelby solution of elastic strain field around a strained inclusion in a homogeneous isotropic elastic solid. To elucidate this deviation, we generated synthetic strain fields using multiple inclusions in two different ways. In the first approach the top $5\%$ strained particles were considered as inclusion centers and the strain on rest of particles in the systems were computed by superposing strain fields due to all inclusions. The second approach was to select particles randomly as inclusion centers and generate synthetic strain fields. The first approach resulted in spatial organisation of inclusion centers that were strongly correlated, whereas the second approach gave rise to spatial organisation that had weak correlations. Further, the investigation of strain correlations of synthetic strain fields revealed that sheared systems show deviation from the $1/r^3$ behavior predicted by Eshelby solution due to strongly correlated formation of inclusions centers. These results are in agreement with the simulation results on foam flows, close to jamming, that had similar conclusions in two dimensions. However, what is interesting is that strain correlations in quiescent glasses decay as $1/r^3$, which is in agreement with Eshelby's predictions \cite{varnik18}. The reason for this disparity between sheared and quiescent system is not clear. A similar analysis for quiescent colloidal glasses might lead to further insights into elasticity and plasticity of amorphous solids.

\section{Acknowledgements}
V.C. acknowledges support from IISER Pune in the form of a startup grant. A.G acknowledges support from Department of Science \& Technology (DST), India for a WOS grant no.SR/WOS-A/PM-34.

\end{document}